\documentclass[prb,draft,twocolumn,superscriptaddress]{revtex4}

\usepackage[final]{graphicx}

\begin{document}



\title{Interface electronic states and boundary conditions for envelope
  functions}

\author{I.~V.~Tokatly}
\email{ilya.tokatly@physik.uni-erlangen.de} 

\affiliation{Moscow Institute of Electronic Technology,
  Zelenograd, 103498, Russia} 
\affiliation{Lerhrstuhl f\"ur Theoretische Festk\"orperphysik,
  Universit\"at Erlangen-N\"urnberg, Staudtstrasse 7/B2, 91054
  Erlangen, Germany}

\author{A.~G.~Tsibizov}
\email{alex@qdn.miee.ru}

\affiliation{Moscow Institute of Electronic Technology,
  Zelenograd, 103498, Russia} 

\author{A.~A.~Gorbatsevich}
\email{aag@qdn.miee.ru}

\affiliation{Moscow Institute of Electronic Technology,
  Zelenograd, 103498, Russia} 

\date{\today}

\begin{abstract} 
The envelope-function method with generalized boundary conditions is
applied to the description of localized and resonant interface states. A
complete set of phenomenological conditions which restrict the
form of connection rules for envelope functions is derived using
the Hermiticity and symmetry requirements. Empirical coefficients in
the connection rules play role of material parameters
which characterize an internal structure of every particular
heterointerface.  
As an illustration we present the derivation of the most general 
connection rules for the one-band effective mass and
4-band Kane models. The conditions for the existence of Tamm-like
localized interface states are established. 
It is shown that a nontrivial form of the
connection rules can also result in the formation of resonant states. 
The most transparent manifestation of such states is the resonant
tunneling through a single-barrier heterostructure.
\end{abstract}


\maketitle


\section{Introduction}

Over 50 years the effective mass or envelope function method is widely used
to describe physical properties of various spatially inhomogeneous
semiconductor systems. Originally the envelope function approach was developed
for external potentials which vary slowly on the atomic scale
\cite{Luttinger}. Nonetheless the application of this method to
semiconductor nanostructures with microscopically abrupt
heterointerfaces is commonly accepted and frequently gives
unexpectedly good results.\cite{Bastard}  

The central question to the effective mass method in the
context of heterostructure applications is how to connect envelopes at
a heterointerface. The simplest and historically the first
connection rules follow from the assumption that the effective
Schr\"odinger equation for the envelope function $\psi({\bf r})$
\begin{equation}
H\psi({\bf r})=E\psi({\bf r}) 
\label{1}
\end{equation}
is valid everywhere in
space, and any  abrupt variation of material parameters can be viewed
as a limit of smooth function. This assumption allows to obtain
boundary conditions by the integration Eq.~(\ref{1}) over an infinitesimally
small distance across the interface \cite{Morrow}. As a result we get
the connection rules \cite{Ben-Daniel,Bastard}
\begin{eqnarray}\nonumber
\psi(+0)&=&\psi(-0),\\ 
{\hat v_{z}}\psi(+0)&=&{\hat v_{z}}\psi(-0),
\quad {\hat v_{z}} = \frac{\partial H}{\partial p_{z}},
\label{SBC}
\end{eqnarray}
($p_{z}$ is the component of momentum perpendicular to
the interface) which are commonly called the standard
or BenDaniel-Duke boundary conditions. For the one-band
effective mass model the boundary conditions Eq.~(\ref{SBC})
are reduced to the continuity of $\psi$ and
$\frac{1}{m}\partial_{z}\psi$, where $m$ is the effective mass.
Obviously, the standard connection rules cannot be universal
since they contain only bulk parameters of materials which constitute
the heterojunction and thus completely neglect internal properties of the
interface. In fact, they work quite well for GaAs/AlGaAs heteropair,
but fail to describe properly more complicated situations.
\cite{Ando0,Laikhtman} The failure of the standard boundary conditions
was also explicitly demonstrated for a number of particular
microscopic models of heterostructures. 
\cite{Sokolov,Romanov,Trzeciakowski2,Grinberg} 

A natural
phenomenological way to take into account the above mentioned results is to
relax the assumption of applicability of Eq.~(\ref{1}) near an
interface and to allow for the discontinuity of both envelops and
their first derivatives. Due to the superposition principle, wave
functions at opposite sides of the interface must be connected by a
linear relation. Assuming locality of this relation we arrive at the
following connection rules 
\begin{equation}
\left( \begin{array}{c}
      \psi(+0)\\ \partial_{z}\psi(+0)
       \end{array}\right) = 
\left( \begin{array}{cc} 
       T_{11} & T_{12}\\T_{21} & T_{22}  
       \end{array}\right)
\left( \begin{array}{c}
      \psi(-0)\\ \partial_{z}\psi(-0)
       \end{array}\right)
\label{T}
\end{equation}
where $T_{ij}\equiv \hat{T}$ is the transfer matrix. 
Boundary conditions of the form Eq.~(\ref{T}) were introduced by Ando
and Mori for the one-band effective mass model.\cite{Ando0} Later
they were also adopted to different milti-band models which describe 
$\Gamma$ and $X$ conduction bands,\cite{Cuypers1,Cuypers2}
degenerate heavy hole and light hole valence bands,\cite{Ando2}
and conduction and valence bands within the spherical
Kane approximation.\cite{Kisin}  

The transfer matrix approach unifies all possible
boundary conditions which have been suggested in the
literature (for particular examples see
Refs.~\onlinecite{Sokolov,Harrison,Kroemer2,Balian,Ivchenko,Einevoll}). 
In fact, the standard boundary conditions Eq.~(\ref{SBC}) are described
by the diagonal transfer matrix: $T_{11}=1$, $T_{12}=T_{21}=0$ and
$T_{22}=m(-0)/m(+0)$;  ${\hat T}$-matrix with $T_{11}=1$, $T_{12}=0$ and 
$T_{21}\ne 0$ corresponds to the introduction of a $\delta$-function 
interface potential;\cite{Sokolov,Kroemer2,Balian,Ivchenko} if only
off-diagonal elements contribute to the transfer matrix we obtain
``inverted'' boundary conditions \cite{Einevoll} (see also
Ref.~\onlinecite{Ando0}) which hold with a high accuracy for GaSb/InAs
interface; \cite{Ando0}  etc. In general all components of the
transfer matrix ${\hat T}$ can be nonzero. They reflect internal
structure of the heterointerface and cannot be expressed in terms of
only bulk parameters. For different particular
cases components of ${\hat T}$-matrix were calculated using empirical
tight-binding and/or 
pseudopotential approaches.\cite{Ando1,Ando2,Cuypers1,Cuypers2}

The general connection rules Eq.~(\ref{T})
can be rigorously justified within the generalized effective mass theory.
\cite{Burt1,Burt2,Foreman2,Foreman3,Takhtamirov} This approach leads
to a set of integral-differential equations for envelope 
functions which are defined using
a single Bloch basis for the whole structure. Coefficients in these
equations and, therefore, envelopes are smooth and continuous
functions even for a system with microscopically abrupt
interfaces. Near an interface the coefficients depend on microscopic
details of the interface. Normally a perturbation, which is 
caused by the interface, is localized at the atomic scale.
Hence, if we are interested in behavior of envelop
functions on the scale which is larger than the lattice constant, we
can use extrapolated bulk envelopes instead of the exact
envelop functions. Despite the exact envelopes
are smooth and continuous, the extrapolated functions obey general
connection rules Eq.~(\ref{T}) with parameters which depend on details
of the interface.\cite{Foreman4} The regular calculation of the
transfer matrix using the generalized effective-mass theory
\cite{Burt1,Burt2,Foreman2,Foreman4} is an extremely tedious
task. However this theory can be considered as a foundation for the
phenomenological introduction of the transfer matrix. 

In this paper we follow such a phenomenological approach and develop a
general method for construction of the transfer matrix
(Sec.~II). Namely, we assume 
that the differential equation (or system of equations) Eq.~(\ref{1})
with piecewise smooth coefficients is not applicable at interface
points (which are the points of discontinuity of the coefficients in
Eq.~(\ref{1})). The Hamiltonian $H$ in Eq.~(\ref{1}) is defined on
a space of 
piecewise smooth and continuous functions with linear connection rules
at the points of discontinuity. We show that Hermiticity of the
Hamiltonian on this space of functions impose the first set of
restrictions on the form of connection rules. A particular
consequence of these restrictions is the conservation of the flow at the
interface. Similar methodology, which can be found in quantum
mechanics text books,\cite{Sudbery} has been recently applied to 
the one-band effective mass model with a general form of the kinetic energy
operator.\cite{Pistol} The second set of
restrictions follows from the symmetry - the transfer matrix must be
invariant with respect to transformations of the interface symmetry
group. These two sets of restrictions strongly reduce the
number of components in the transfer matrix. The rest of 
${\hat T}$-matrix along with band offsets should be considered as
empirical parameters which are defined from experiment (see for example
Ref.~\onlinecite{Trzeciakowski1}) and/or {\it ab initio}
calculations. The method developed in Sec.~II
is closely related to the common method of invariants\cite{Bir} 
which allows to construct effective ${\bf k\cdot p}$ Hamiltonians for bulk
semiconductors using only Hermiticity and symmetry requirements. 

The general connection rules Eq.~(\ref{T}) allow to describe various
physical consequences of nontrivial internal details of a
heterointerface. It has been demonstrated in
Ref.~\onlinecite{Trzeciakowski1} that the use of general boundary
conditions removes quantitative discrepancies between square well
calculations and experiment.\cite{Nelson} There are also more transparent
qualitative effects which come from the complex structure of
boundary conditions. For example, in multi-band models the off-diagonal
element $T_{21}$ (which is equivalent to an interface
$\delta$-potential) is responsible for the interface heavy-light 
hole\cite{Ivchenko} and $\Gamma-X$ electron \cite{Cuypers2,Ando3,Fu}
mixing. In the present paper we concentrate on a description of
interface localized and resonant states.

In 1932 Tamm\cite{Tamm} demonstrated the existence of electronic
states localized at a surface of a semiconductor. It is quite natural
to expect that similar localized states with energies inside the
forbidden gap can occur at an abrupt heterojunction. Such a possibility
was qualitatively considered by James \cite{James} and later by Zhu
and Kroemer.\cite{Kroemer2} In Ref.~\onlinecite{Kroemer1} the
existence of interface donor states was postulated to explain
anomalous transport properties of undoped InAs/AlSb
quantum-well structures. However, general conditions for occurrence of
interface states remained unclear for a long time. This problem was
addressed in Ref.~\onlinecite{GT} and recently in
Ref.~\onlinecite{Kolesnikov} using the tight-binding approach. Since
the effective-mass method is of extreme importance for heterostructure
applications, it is desirable to have a description of
interface states in terms of envelope functions.

It is known that envelope-function models with the
standard boundary conditions possess interface states.\cite{Suris,Raichev}
However the corresponding energy levels always lie in the region of band
offsets outside the energy gap of a heterojunction.\cite{Suris,Raichev}
An interesting exception is the formation of localized states at
heterojunctions with band inversion.\cite{Suris,Volkov,Sham}  These
states have a topological nature and are related to the supersymmetry of an
inverse contact.\cite{Volkov}

In Sec.~III of our paper we show that different types of interface
states which have energies in the forbidden gap can be described using
generalized connection rules. We apply the boundary conditions
derived in Sec.~II to the one-band effective-mass model and
to the 4-band Kane model which describe $\Gamma$-point states in
III-V semiconductors. We derive general conditions for the
existence of interface states, and discuss the physical
meaning of the off-diagonal components in the transfer matrix. In
Sec.~IV we study a scattering problem and demonstrate the existence of
resonant states and resonant tunneling through a single barrier
stricture, which are related to nonzero off-diagonal elements in the
transfer matrix. In Sec.~V we summarize our results. 

\section{General approach and basic equations}

To establish a general form of connection rules we consider the standard
statement of a problem within the envelope-function approach that is to
find eigen functions $\psi({\bf r})$ and eigen values $E$ of a
Hamilton operator $H$.
Let, as usual, the Hamiltonian $H$ be a second-order matrix
differential operator
\begin{equation}
H=H_0({\bf r})-
\frac{1}{2}M_{\alpha\beta}\partial_{\alpha}\partial_{\beta}+
iL_{\alpha}\partial_{\alpha}
\label{2}
\end{equation}
where $H_0({\bf r})$, $M_{\alpha\beta}$ and $L_{\alpha}$ are $m\times
m$ Hermitian 
matrixes and $\alpha,\beta=x,y,z$. We assume that the
growth direction of a structure coincides with $z$-axis and the system
is spatially homogeneous in $x-y$-plane.  In this case
wave functions take the form $\psi({\bf r})=e^{i{\bf
k_{\perp}r}}\psi(z)$ ($\bf k_{\perp}$ is the momentum
perpendicular to $z$-axis). The function $\psi(z)$ is a solution to the
one-dimensional Schr\"odinger equation with the Hamiltonian
\begin{equation}
H_z=h(z)-\gamma\partial_z^2 +iP\partial_z,
\label{3}
\end{equation}
where
\begin{eqnarray} \nonumber
h(z)&=&H_0({\bf r})-
\frac{1}{2}M_{\alpha\beta}k_{\perp \alpha}k_{\perp \beta}+
iL_{\alpha} k_{\perp \alpha},\\
\gamma&=&M_{zz},\\ \nonumber P&=& L_z +iM_{\alpha z}k_{\perp\alpha}
\label{4}
\end{eqnarray}
We consider the system which
consists of $N$ regions with different material parameters and
introduce notations  
$z_{n-1}$ and $z_n$ for the left and the right boundaries of the $n$th
region ($n=1...N$). Assume that $h(z)$ is a piecewise
smooth function of $z$, whereas $\gamma$ 
and $P$ are piecewise constant functions. In the $n$th region
($z_{n-1}<z<z_n$) matrixes $h(z)$, 
$\gamma$ and $P$ respectively take the values $h_n(z)$, $\gamma_n$
and $P_n$, where $h_n(z)$ is a smooth and continuous function, and 
$\gamma_n$ and $P_n$ are constants.

Let ${\cal W}_{T}$ be the space of functions $\psi(z)$ which are piecewise
smooth and 
square integrable on every interval $z_{n-1}<z<z_n$. Besides, at every point
$z_{n}$ values $\psi(z_n+0)$, $\partial_z \psi(z_n+0)$ and $\psi(z_n-0)$,
$\partial_z \psi(z_n-0)$ are connected by linear relations
\begin{equation}
\left( \begin{array}{c}
\psi(z_n+0)\\ \partial_z \psi(z_n+0)
\end{array} \right)= T_n\left( \begin{array}{c}
\psi(z_n-0)\\ \partial_z \psi(z_n-0)\end{array} \right),
\label{5}
\end{equation}
where $T_n$ ($n=1..N$) are $2m\times 2m$ matrixes. 

The differential operator $H_{z}$ Eq.~(\ref{3}) is well defined on the
space ${\cal W}_{T}$, but not necessarily Hermitian. The Hermiticity
condition impose a restriction on a possible form of the transfer matrixes
$T_{n}$. 

By the definition the operator $H_{z}$ is Hermitian on the space
${\cal W}_{T}$ if 
\begin{equation}
I=<\varphi|H_{z}|\psi>-<\psi|H_{z}|\varphi>^*=0,
\label{6}
\end{equation}
where $\varphi$ and $\psi$ belong to ${\cal W}_{T}$. Matrix
elements in Eq.~(\ref{6}) are defined as integrals over the whole
structure with the points of discontinuity being excluded
\begin{equation}
<\varphi|H|\psi>= \sum_{n=1}^N \int\limits_{z_{n-1}}^{z_n}\varphi^+
(h_n(z)-\gamma_n\partial_z^2 +iP_n\partial_z)\psi dz.
\label{7}
\end{equation}
We assume for definiteness that $z_0=-\infty$, $z_N=\infty$ and
$\psi(\pm\infty)=0$. It is convenient to introduce $2m$-component
vectors
\begin{equation}
\Psi=\left( \begin{array}{c}
\psi(z)\\ \partial_z \psi(z)
\end{array} \right), \qquad \Phi=\left( \begin{array}{c}
\varphi(z)\\ \partial_z \varphi(z)\end{array} \right)
\label{8}
\end{equation}
and a "current" operator $J_n$ which acts on these vectors
\begin{equation}
J_n=\left[ \begin{array}{cc}
P_n & -i\gamma_n\\ i\gamma_n & 0
\end{array} \right].
\label{9}
\end{equation}
Integration by parts in Eq.~(\ref{6}) leads to the following
expression  for the quantity $I$ 
\begin{equation}
I=i\sum_{n=1}^N \Delta_n\{\Phi^+J\Psi\},
\label{10}
\end{equation}
where $\Delta_n\{\Phi^+J\Psi\}$ is a jump of the quantity
$\Phi^+J\Psi$ at the point $z=z_n$: 
\begin{eqnarray}\nonumber
\Delta_n\{\Phi^+J\Psi\} &=& \Phi^+(z_{n}+0)J\Psi(z_{n}+0)\\ \nonumber 
                        &-& \Phi^+(z_{n}-0)J\Psi(z_{n}-0).
\end{eqnarray}
The Hermiticity condition $I=0$ is fulfilled if
$$
\Delta_n\{\Phi^+J\Psi\}=0
$$
for every boundary and any pair of functions $\psi$ and $\varphi$
from ${\cal W}_{T}$. Using the definition of transfer matrixes
Eq.~(\ref{5}) we arrive at the Hermiticity condition of the following form
\begin{equation}
J_n=T^+_nJ_{n+1}T_n, 
\label{11}
\end{equation}
which means the invariance of the "current"
operator $J$ under the transfer across a discontinuity point.

To simplify formulas we consider bellow a system
with a single boundary at $z=0$ which separates left ($n=1=L$)
and right ($n=2=R$) regions. Hence the connection rules
Eq.~(\ref{5}) take the form 
\begin{equation}
\Psi_R(0)=T\Psi_L(0), \label{BC}
\end{equation}
where the transfer matrix $T$ must satisfy the following Hermiticity condition
\begin{equation}
J_r=T^+J_l T. \label{herm}
\end{equation}

Another set of restrictions follows from the fact that
$T$-matrix in Eq.~(\ref{BC}) should be invariant with respect to the
symmetry group $G$ of the interface plane:
$$
\hat D(g)T\hat D^{-1}(g)=T,
$$
where $\hat D(g)$ is $2m\times 2m$ matrix which corresponds to an
element $g$ of the group $G$. Since $\psi(0)$ and $\partial_z\psi(0)$
have the same  
transformation properties with respect to operations of the
interface symmetry group $G$, matrixes $\hat D(g)$ take a block
diagonal form
\begin{equation}
\hat D(g)=\left[ \begin{array}{cc}
D(g) & 0\\ 0& D(g)
\end{array} \right],
\label{14}
\end{equation}
where $m\times m$ matrixes $D(g)$ form a representation (reducible
in general case) of the group $G$ in the basis which corresponds to the
bulk Hamiltonian Eq.~(\ref{2}). Hence the symmetry requirements can be
written independently for every $m\times m$ block $T_{ij}$ ($i,j=1,2$)
of the full transfer matrix $T$: 
\begin{equation}
D(g)T_{ij}D^{-1}(g)=T_{ij}. \label{sim}
\end{equation}

Equations (\ref{herm}) and (\ref{sim}) provide a complete set of
phenomenological requirements which restrict the form of the general
connection rules Eq.~(\ref{BC}). In the next section we present a
solution of these equations for the one-band effective mass model and
4-band Kane model.

\section{Localized Tamm-like interface states within the envelope
function approach}

\subsection{One-band effective mass approximation}
In this section we consider a single heterointerface located at the point
$z=0$. The corresponding band diagram is shown in Fig.~1. 
\begin{figure}
  \includegraphics[width=0.4\textwidth]{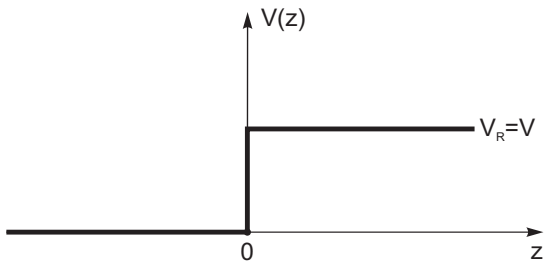}
  \caption{One-band potential profile of the heterojunction (see text).}
  \label{fig1}
\end{figure}
The one-band
effective mass Hamiltonian takes the form 
\begin{equation}
H_n=-\frac{\nabla^2}{2m_n}+V_n,
\label{16}
\end{equation}
where $n=L,R$ and $m_{L,R}$ and $V_{L,R}$ are effective masses and
potentials in the left ($L$) and the right ($R$) bulk regions
respectively. The
"current" operator Eq.~(\ref{9}) for this model is $J_n=\sigma_y/2m_n$ 
($\sigma_{y}$ is the Pauli matrix). Hence
the Hermiticity condition Eq.~(\ref{herm}) takes the form
\begin{equation}
\beta \sigma_y=T^+\sigma_yT, \qquad \beta=\frac{m_R}{m_L}.
\label{17}
\end{equation}
The general solution to this equation is
\begin{equation}
T=t e^{i\chi}\sqrt{\beta},
\label{18}
\end{equation}
where $\chi$ is an arbitrary phase and $t$ is a real $2\times 2$
matrix with unit determinant: 
\begin{equation}
\det t = t_{11}t_{22}-t_{12}t_{21} =1.
\label{19}
\end{equation}

Let us study localized solutions which are allowed by the boundary 
condition Eq.~(\ref{BC}) with the transfer matrix Eq.~(\ref{18}). 
We assume for definiteness that $V_L=0$, and $V_R=V>0$ is a band
offset (see Fig.~1). Wave functions of interface states take the
following general form 
\begin{eqnarray} \nonumber
\psi_R &=& Ae^{i{\bf k_\perp r}}e^{-\kappa_Rz}, \qquad z>0 \\
\psi_L &=& Be^{i{\bf k_\perp r}}e^{\kappa_Lz}, \qquad z<0.
\label{20}
\end{eqnarray}
The energy of the localized state is defined as
$$
E=-\frac{\kappa_L^2}{2m_L}+\frac{{\bf k_\perp}^2}{2m_L},
$$
where $\kappa_L^2/2m_L$ is the binding energy. Quantity $\kappa_R$
in Eq.~(\ref{20}) is related to $\kappa_{L}$ by the equation
\begin{equation}
\kappa_R=\sqrt{\beta\kappa_L^2+q^2}, \qquad q^2=2m_RV+
(1-\beta){\bf k_\perp}^2.
\label{21}
\end{equation}
Parameter $q$ in Eq.~(\ref{21}) describes a degree of the heterointerface
asymmetry which is related both to the band offset and to the difference
of effective masses.

Substitution of Eq.~(\ref{20}) into the boundary conditions leads to
the following dispersion equation for $\kappa_L$
\begin{equation}
\sqrt{\beta\kappa_L^2+q^2}=-\frac{t_{21} + t_{22}\kappa_L}{t_{11} +
t_{12}\kappa_L}.
\label{22}
\end{equation}

It is natural to assume that the diagonal elements of the matrix $t$
are positive ($t_{11}>0$, $t_{22}>0$). In this case 
Eq.~(\ref{22}) has real solutions only if $t_{21}$ and/or $t_{12}$ are
negative. There are two types of solutions. Solutions of the first
type correspond to the case $t_{21}<0$, $t_{12}\ge 0$. The existence
condition takes the form 
\begin{equation}
-\frac{t_{21}}{t_{11}}>q.
\label{23}
\end{equation}
These states occur near the lowest band edge and move down into the
energy gap with increase of $|t_{21}|$.
Equation (\ref{23}) shows that the existence of such solutions is
restricted by the value of the asymmetry parameter $q$
Eq.~(\ref{21}). It was mentioned in the Introduction that a nonzero
element $t_{21}$ can be modeled by an interface
$\delta$-potential. Negative $t_{21}$ corresponds to an attractive
interface potential. Hence Eq.~(\ref{23})
is analogous to the well known condition for the existence of bound
states in a potential well with asymmetric barriers.\cite{Landau}
If $m_{L}\ne m_{R}$ ($\beta\ne 1$) the asymmetry parameter $q$ depends
on 2D momentum $k_\perp$ (see Eq.~(\ref{21})). Therefore Eq.~(\ref{23})
defines a line in ${\bf k}_\perp$-space which separates localized 2D
interface states and delocalized 3D continuum states. 
Analogous 2D-3D transformations were recently studied in asymmetric quantum
wells \cite{Kopaev}. 

Solutions of the second type are related to negative values of the
second off-diagonal element $t_{12}$. They exist at arbitrary 
$t_{12}<0$ and $t_{21}\ge 0$. At small negative $t_{12}$ the
energy levels, which correspond to these states, lie
deep in the forbidden gap and approach
the band edge with increase of $|t_{12}|$. The solutions of the second
type can be naturally viewed as states which originate from a
lower (for example valence) band and move up with increase of an
effective interface potential. In Sec.IIIC we shall return to this
point and discuss a possible interpretation of the element $t_{12}$ in
terms of a local perturbation of remote bands.  

Obviously, the solutions of both types can coexist if both
off-diagonal elements are negative.

\subsection{Interface states in the 4-band Kane model}

To study interface states in multi-band systems and to
illustrate the importance of the symmetry requirements Eq.~(\ref{sim}) we
consider the 4-band Kane model which describes $\Gamma$-point states in
III-V zinc blend semiconductors without spin-orbit splitting. We
consider a single $(001)$ heterojunction and find
the interface states which originate from $\Delta$-line states with
${\bf k_\perp}=0$.  

It is convenient to introduce the following basis
\begin{equation}
\{\chi_j\}=\{|S\rangle,|Z\rangle,|X_+\rangle,|X_-\rangle\},
\label{24}
\end{equation}
where $|X_{\pm}\rangle=\frac{1}{\sqrt{2}}(|X\rangle\pm
|Y\rangle)$. In this basis the $\Delta$-line 
Hamiltonian takes a block diagonal form \cite{Bir}
\begin{equation}
H_n=\left[ \begin{array}{cc}
H^n_{sz} & 0\\ 0 &H^n_{\pm}
\end{array} \right],
\label{25}
\end{equation}
where $H^n_{sz}$ and $H^n_{\pm}$ are $2\times 2$ matrixes which
correspond to $|S\rangle$, $|Z\rangle$ and $|X_{+}\rangle$,
$|X_{-}\rangle$ pairs of states respectively ($n=L,R$)
\begin{eqnarray}\nonumber
H^n_{sz}&=&\left[ \begin{array}{cc}
E_{cn}-\frac{\partial_z^2}{2m_n} &P_n\partial_z\\
-P_n\partial_z &E_{vn}-\frac{\partial_z^2}{2m'_n}
\end{array} \right], \\
H^n_{\pm}&=&
\left(E_{vn}+\frac{\partial_z^2}{2m_{vn}}\right)I.
\label{26}
\end{eqnarray}
In Eq.~(\ref{26}) $E_{cn}$ and $E_{vn}$ are the energies of conduction
and valence band edges respectively and $I$ is the $2\times 2$ unit matrix.

First we establish the general form of boundary conditions for this
system. The transfer matrix 
\begin{equation}
T=\left[ \begin{array}{cc}
\hat T_{11} &\hat T_{12} \\ \hat T_{21} &\hat T_{22}
\end{array} \right],
\label{TKane}
\end{equation}
which enters the connection rules Eq.~(\ref{BC}) consists of four
$4\times 4$ blocks. Every block $\hat{T}_{ij}$
($i,j=1,2$) must satisfy the symmetry conditions Eq.~(\ref{sim}). 
The symmetry group of
$(001)$ plane for zinc blend structure is the group
$C_{2v}$ \cite{Ivchenko} which has four elements: the unit element
$E$, a second-order axis $C_2$ and two mutually perpendicular
reflection planes $\sigma_1$ and $\sigma_2$. This group has four classes
and thus four irreducible representations. Each function from
the set $\{\chi_j\}$ Eq.~(\ref{24}) 
is the basis function for one of the irreducible representations.
Namely, functions $|S\rangle$ and $|Z\rangle$ correspond to the same
representation $A_1$, whereas functions $|X_+\rangle$ and $|X_-\rangle$
are related to the representations $B_1$ and $B_2$ respectively.
Thus matrixes $D(g)$ which enter the symmetry conditions Eq.~(\ref{sim})
have a diagonal form in the basis Eq.~(\ref{24})
\begin{eqnarray}\nonumber
D(E)&=&\text{diag} (1,1,1,1), \\ \nonumber D(C_2)&=&\text{diag}(1,1,-1,-1),\\
D(\sigma_1)&=&\text{diag}(1,1,1,-1), \\ \nonumber D(\sigma_2)
&=&\text{diag}(1,1,-1,1),
\label{27}
\end{eqnarray}
where $\text{diag}(...)$ stands for the set of diagonal elements. 
Straightforward calculations show that the general solution to
Eq.~(\ref{sim}) with $D(g)$ Eq.~(\ref{27}) takes the form 
\begin{equation}
\hat T_{ij}=\left[ \begin{array}{ccc}
\hat T^{ij}_{sz} & 0 & 0\\ 0 & T^{ij}_+ & 0 \\ 0 & 0 &T^{ij}_-
\end{array} \right],
\label{28}
\end{equation}
where $\hat T^{ij}_{sz}$ is an arbitrary $2\times 2$ matrix and
$T^{ij}_{\pm}$ are arbitrary numbers. Consequently, pairs of 
states $(|S\rangle,|Z\rangle)$ and $(|X_+\rangle,|X_-\rangle)$ as well
as the states $|X_+\rangle$ 
and  $|X_-\rangle$ are decoupled due to the symmetry requirements.
Thus, we get three independent problems. Two of
them correspond to the solution of two independent one-band Schr\"odinger
equations with
general boundary conditions (see Sec.~IIIA) for decoupled $|X_+\rangle$ and
$|X_-\rangle$ valence bands. It is worth to mention 
that if $T^{ij}_+\ne T^{ij}_-$, which is allowed by the symmetry, the
connection rules Eqs.~(\ref{BC}), (\ref{TKane}), (\ref{28}) lead to a
heavy-light hole mixing. In fact, the connection rules
\begin{eqnarray}  \nonumber
\partial_z\psi_{XR}(0)&=&\partial_z\psi_{XL}(0)+T_{l-h}\psi_Y(0),\\
\psi_{X,YL}(0)&=&\psi_{X,YR}(0),
\label{l-h1}
\end{eqnarray}
which are used \cite{Ivchenko} to describe
the heavy-light hole mixing at the normal hole incidence, represent a
particular case of the $T$-matrix Eqs.~(\ref{TKane}), (\ref{28}) with
$T^{jj}_{\pm}=1$, $T^{12}_{\pm}=0$ and $T^{21}_{+}=-T^{21}_{-}=T_{l-h}$.  

To describe localized states which correspond to the
subspace $\{|S\rangle,|Z\rangle\}$ we have to solve the two-band Schr\"odinger
equation with the Hamiltonian $H_{sz}$ Eq.~(\ref{26}). The boundary
conditions to this equation are defined via the transfer matrix 
$\hat T_{sz}$ which has 
no symmetry restrictions since both $|S\rangle$ and $|Z\rangle$ 
correspond to the same representation $A_1$ of the interface
group $C_{2v}$.

To simplify further calculations we neglect the second-derivative terms in
$H_{sz}$. This reduces the problem to the solution of two coupled
first-order differential equations 
\begin{equation}
\left[ \begin{array}{cc}
E_{cn}-E &P_n\partial_z\\
-P_n\partial_z &E_{vn}-E
\end{array} \right]\psi_n(z)=0.
\label{29}
\end{equation}
Since the highest spatial derivative in Eq.~(\ref{29}) is of
the first order we should not include the derivatives of wave functions in
the boundary condition. Therefore the transfer matrix $\hat T_{sz}$
has only one nonzero $2\times 2$ block $T^{11}_{sz}\equiv T_{sz}$
which is restricted only by the Hermiticity condition. The "current"
operator for the problem Eq.~(\ref{29}) takes the form
$J_j=P_j\sigma_y$. Hence the Hermiticity condition
Eq.~(\ref{herm}) formally coincides with Eq.~(\ref{17})
\begin{equation}
\gamma \sigma_y=T^+_{sz}\sigma_yT_{sz}, \qquad \gamma=\frac{P_L}{P_R}.
\label{30}
\end{equation}
The solution to Eq.~(\ref{30}) is an arbitrary real $2\times 2$ matrix with
fixed determinant and with an arbitrary phase (see Sec.~IIIA)
\begin{equation}
T_{sz}=t e^{i\chi}\sqrt{\gamma},\qquad \det t =1.
\label{31}
\end{equation}
Considering a localized interface solution
\begin{equation}
\psi_R(z)=A\left(\begin{array}{c} u_R\\ v_R\end{array}\right)
e^{-\kappa_Rz}, \quad
\psi_L(z)=B\left(\begin{array}{c} u_L\\ v_L\end{array}\right)
e^{\kappa_Lz}
\label{32}
\end{equation}
we get the following dispersion equation
\begin{equation}
F_R(E)=\frac{t_{12} - t_{11}F_L(E)}{t_{22} -
t_{21}F_L(E)},
\label{33}
\end{equation}
where
\begin{equation}
F_n(E)=\sqrt{\frac{E-E_{vn}}{E_{cn}-E}}.
\label{34}
\end{equation}
The localized solution of the form Eq.~(\ref{32}) exists if the energy
$E$ lies in the forbidden gap of the heterojunction
$$
\max\{E_{vL},E_{vR}\}<E<\min\{E_{cL},E_{cR}\}
$$
which is shown by shaded region in Fig.~2.
\begin{figure}
  \includegraphics[width=0.45\textwidth]{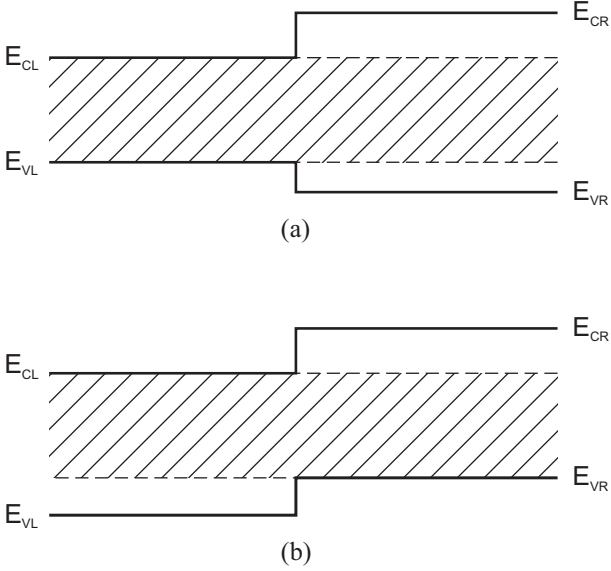}
  \caption{(a) Band digram for a heterojunction of the I type; (b)
    Band digram for a heterojunction of the II type }
  \label{fig2}
\end{figure}
It is convenient to introduce a new variable
\begin{equation}
x(E)=F_L(E)=\sqrt{\frac{E-E_{vL}}{E_{cL}-E}}
\label{35}
\end{equation}
and rewrite the dispersion equation Eq.~(\ref{33}) in the form
\begin{equation}\label{36}
F_{R}(x) = \frac{t_{12} - t_{11}x}{t_{22} - t_{21}x},
\end{equation}
where the function $F_{R}(x)$ is defined as follows
\begin{equation}
F_{R}(x)=\sqrt{
\frac{E_{vL}-E_{vR}+x^2(E_{cL}-E_{vR})}{E_{cR}-E_{vL} +
  x^2(E_{cR}-E_{cL})}}.
\label{35a}
\end{equation}
We analyze the solutions
of Eq.~(\ref{36}) under the natural assumption $t_{11}>0$ and  $t_{22}>0$.
We also assume that $E_{cR}>E_{cL}$ i.e. the conduction band offset 
$V=E_{cR}-E_{cL}$ is positive (see Fig~2). Therefore a heterointerface
of the I type (Fig.1a) corresponds to the condition
\begin{equation}
E_{vR}<E_{vL},
\label{37}
\end{equation}
whereas the inverse inequality 
\begin{equation}
E_{vR}>E_{vL}
\label{38}
\end{equation}
holds if the heterostructure belongs to the II type (see Fig.~2b).

First we consider heterointerfaces of the I type. In this case
the region of the energy gap (shaded region in Fig.~2a)
$$E_{vL}<E<E_{cL}$$
maps to the region 
$$0<x<\infty$$ 
of the variable $x$. When $x$ goes from $0$ to $\infty$, the function
$F_R(x)$ in the  right hand side of Eq.~(\ref{36}) monotonically
increases from
\begin{equation}
F_{\min}^{I}=\sqrt{(E_{vL}-E_{vR})/(E_{cR}-E_{vL})} 
\label{39.1}
\end{equation}
at $x=0$ (which corresponds to $E=E_{vL}$) to the value 
\begin{equation}
F_{\max}^{I}=\sqrt{(E_{cL}-E_{vR})/(E_{cR}-E_{cL})}
\label{39.2}
\end{equation} 
at $x=\infty$ ($E=E_{cL}$). The function $F_{R}(x)$ for a
heterojunction of the I type is shown by solid line in Fig.~3. 
\begin{figure}
  \includegraphics[width=0.4\textwidth]{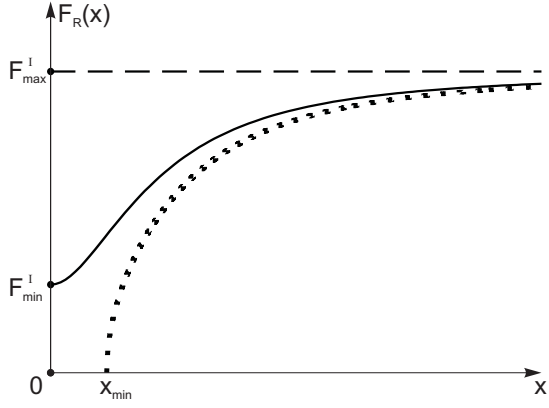}
  \caption{Dependence $F_{R}$ vs $x$ (see Eq.~(40) in the text). Solid (dotted)
  line shows $F_{R}(x)$ for the heterojunction of the I (II) type.}
  \label{fig3}
\end{figure}
The
behavior of the right hand side in Eq.~(\ref{36}) depends on the signs of
the off-diagonal elements $t_{12}$ and $t_{21}$. The dispersion
equation Eq.~(\ref{36}) have real solutions if at least one of the
off-diagonal elements is positive. If $t_{12}$ and $t_{21}$ have
opposite signs there exists only one solution to Eq.~(\ref{36}).

Let us analyse different cases separately.

(i) $t_{12}>0$, $t_{21}\le 0$. In this case the right hand side in
Eq.~(\ref{36}) is a decreasing function of $x$. A solution exists if 
\begin{equation}
\frac{t_{12}}{t_{22}} > \sqrt{\frac{E_{vL}-E_{vR}}{E_{cR}-E_{vL}}}=
F_{\min}^{I}.
\label{39}
\end{equation}
This solution can be identified with an acceptor interface state which
originates from the valence band and move up from the edge of the
valence band $E_{vL}$ to the edge of the conduction band $E_{cL}$ with
increase of $t_{12}$.

(ii) $t_{21}>0$, $t_{12}\le 0$. A solution to Eq.~(\ref{36}) exists
under the following condition 
\begin{equation}
\frac{t_{21}}{t_{11}} > \sqrt{\frac{E_{cR}-E_{cL}}{E_{cL}-E_{vR}}}=
\left(F_{\max}^{I}\right)^{-1}.
\label{40}
\end{equation}
This is a donor interface state originated from the conduction
band. With increase of $t_{21}$ the energy level, which corresponds to
this state, moves through the energy gap from $E_{cL}$ to $E_{vL}$.

If the heterointerface belongs to the II type, then $E_{vR}$ forms a
lower bound of the energy gap (shaded region in Fig.~2b). Hence the
energy gap
$$E_{vR}<E<E_{cL}$$
maps to the region
$$ 
x_{\min}\equiv\sqrt{(E_{vR}-E_{vL})/(E_{cL}-E_{vR})}<x<\infty
$$
in the $x$-axis.

The function $F_R(x)$ in Eq.~(\ref{36}) increases from zero at $x =x_{\min}$
($E=E_{vR}$) to the value $F_{\max}^{II}=F_{\max}^{I}$
Eq.~(\ref{39.2}) at $x=\infty$ ($E=E_{cL}$). This function is shown by
the dotted line in Fig.~3. Consequently, only the
condition for  the existence of the acceptor state is changed. For the
heterointerface of the II type the  existence condition Eq.~(\ref{39}) is
replaced by the inequality 
\begin{equation}
\frac{t_{12}}{t_{11}} > \sqrt{\frac{E_{vR}-E_{vL}}{E_{cL}-E_{vR}}} =
x_{\min}.
\label{41}
\end{equation}
The condition for the existence of the donor state Eq.~(\ref{40})
remains unchanged. 

Acceptor and donor interface states coexist if
$t_{12}t_{21}<t_{11}t_{22}$ ($t_{12}>0$, $t_{21}>0$) and the
conditions Eq.~(\ref{39}) (Eq.~(\ref{41}) for the II-type structure)
and Eq.~(\ref{40}) are fulfilled simultaneously.

\subsection{Physical meaning of the off-diagonal elements in the
  transfer matrix}

In this subsection we discuss a possible interpretation of the elements
$T_{12}$ and $T_{21}$ in the transfer matrix for the one-band effective
mass model. 

It is well known\cite{Ando1,Balian} that the element $T_{21}$
can be interpreted as an interface $\delta$-function potential. Indeed,
a nonzero $T_{21}$ introduces a jump of the first derivative
which is proportional to the value of the wave function, exactly as
an interface $\delta$-potential does. The physical meaning of the
second element $T_{12}$ is less clear (see, for example, discussions
in Refs.~\onlinecite{Ando1,Balian,Foreman4}). In Sec.~IIIA we have
shown that there exist localized states which are related to
nonzero $T_{12}$. These states behave as ``acceptor'' states which
originate from remote lower bands. Therefore, it is natural to expect
that the element $T_{12}$ is related to a local perturbation of these
remote bands though they are not explicitly included in the bulk
one-band Hamiltonian. To confirm this interpretation we consider a
one-band model as the limit of the two-band model with the following
Hamiltonian
\begin{equation}
H = \left[ \begin{array}{cc}
E_{c}(z) &P\partial_z\\
-P\partial_z &E_{v}(z)
\end{array} \right] + \left[ \begin{array}{cc}
                             g_{c} & 0\\
                               0   & g_{v}\end{array} \right]\delta(z),
\label{H2band}
\end{equation}
where $E_{c(v)}(z)$ describes the profile of the conduction (valence)
band. The first term in Eq.~(\ref{H2band}) corresponds to bulk
regions whereas the second term models a local perturbation which is
caused by the interface.

The boundary conditions for the two-component envelope function can be
obtained by the integration of the Schr\"odinger equation
\begin{equation}
H\left( \begin{array}{c} \psi \\ \varphi \end{array} \right) =
E\left( \begin{array}{c} \psi \\ \varphi \end{array} \right)
\label{2band-eq}
\end{equation}
over a small segment $[-a,a]$ and taking the limit $a\rightarrow 0$.  
The result of the integration takes the form
\begin{eqnarray}\nonumber
\psi_{R}(0)-\psi_{L}(0) &=& \frac{g_{v}}{P}\lim_{a\rightarrow 0}
                        \int_{-a}^{a}\delta(z)\varphi(z)dz, \\ \nonumber
\varphi_{R}(0)-\varphi_{L}(0) &=& - \frac{g_{c}}{P}\lim_{a\rightarrow 0}
                          \int_{-a}^{a}\delta(z)\psi(z)dz.
\end{eqnarray}
Using the identity
$$
\int^{a}_{-a}\delta(z)\theta(z)dz = \frac{1}{2}
$$
($\theta(z)$ is the Heaviside function) we arrive at the 
connection rules
\begin{equation}
\left( \begin{array}{c}
      \psi_{R}\\ \varphi_{R}
       \end{array}\right) = 
\left( \begin{array}{cc} 
       t_{11} & t_{12}\\t_{21} & t_{22}  
       \end{array}\right)
\left( \begin{array}{c}
      \psi_{L}\\ \varphi_{L}
       \end{array}\right),
\label{2bandBC}
\end{equation}
where
\begin{eqnarray}\label{t2band1}
& & t_{11} = t_{22} = \frac{1-g_{c}g_{v}/4P^{2}}{1+g_{c}g_{v}/4P^{2}},
\\ \label{t2band2}
t_{12} &=& \frac{g_{v}/P}{1+g_{c}g_{v}/4P^{2}} , \quad
t_{21} = - \frac{g_{c}/P}{1+g_{c}g_{v}/4P^{2}}.
\end{eqnarray}
Equations (\ref{2bandBC})-(\ref{t2band2}) show that both off-diagonal
elements for two-band model are reproduced by the simple interface term
Eq.~(\ref{H2band}), though the transfer matrix in Eq.~(\ref{2bandBC})
is still not of the most general form (compare to Eq.~(\ref{31})). We
can also clarify the physical meaning of the interface solutions for the
two-band model which has been considered in Sec.~IIIB. According to
the results of Sec.~IIIB the interface solution of acceptor type
exists if $t_{12}$ is positive. Positive $t_{12}$ corresponds to
positive $g_{v}$ (see Eq.~(\ref{t2band2})) and, consequently, to a
local perturbation of the valence band which is attractive for
holes. Analogously, the solution of the donor type is related to a
positive value of $t_{21}$ which corresponds to  negative $g_{c}$
and a local perturbation of the conduction
band, attractive for electrons. 

Let us derive the one-band model which is related to the two-band
Hamiltonian Eq.~(\ref{H2band}). If the energy $E$ in
Eq.~(\ref{2band-eq}) is close to the edge of the conduction band
$$
|E - E_{c}(z)|/\Delta(z)\ll 1
$$  
($2\Delta = E_{c}-E_{v}$ is the energy gap), we can express the lower
component of the spinor in Eq.~(\ref{2band-eq}) in terms of the upper
component 
\begin{equation}
\varphi(z) \approx -\frac{P}{2\Delta(z)}\partial_{z}\psi(z).
\label{phi}
\end{equation}
The upper component $\psi(z)$ plays the role of the wave function which in
the bulk regions satisfies the one-band Schr\"odinger equation
\begin{equation}
- \partial_{z}\frac{1}{2m}\partial_{z} \psi(z) =
(E - E_{c}(z))\psi(z),
\label{1band-eq}
\end{equation}
where $m(z)=\Delta(z)/P^{2}$ is the effective mass. Equation
(\ref{1band-eq}) should be supplemented by boundary conditions which
are obtained by the substitution Eq.~(\ref{phi}) to the connection
rules Eq.~(\ref{2bandBC}). The final boundary conditions for the
one-band model take the form
\begin{equation}
\left( \begin{array}{c}
      \psi_{R}\\ \partial_{z}\psi_{R}
       \end{array}\right) = 
\left( \begin{array}{cc} 
       T_{11} & T_{12}\\T_{21} & T_{22}  
       \end{array}\right)
\left( \begin{array}{c}
      \psi_{L}\\ \partial_{z}\psi_{L}
       \end{array}\right),
\label{1bandBC}
\end{equation}
with the following elements of the transfer matrix
\begin{eqnarray}\label{t1band1}
& & T_{11} =\frac{m_{L}}{m_{R}}T_{22} = 
\frac{1-g_{c}g_{v}/4P^{2}}{1+g_{c}g_{v}/4P^{2}},
\\ \label{t1band2}
T_{12} &=& - \frac{g_{v}/2\Delta_{L}}{1+g_{c}g_{v}/4P^{2}} , \quad
T_{21} = \frac{2m_{R} g_{c}}{1+g_{c}g_{v}/4P^{2}}.
\end{eqnarray}
Thus, the off-diagonal element $T_{12}$ Eq.~(\ref{t1band2}) is
proportional to the strength $g_{v}$ of the local perturbation of the remote
valence band. A potential, which is attractive for holes,
corresponds to positive $g_{v}$ and thus negative $T_{12}$. This
explains the 
results of Sec.~IIIA and confirms our interpretation of the interface state 
related to the element $T_{12}$. 

\section{Resonant tunneling through a single barrier with complex interfaces}

As we have seen in Sec.~III, a nontrivial internal structure of a single
heterointerface which is described by the generalized boundary
conditions allows for the existence of localized interface
states. In this section we show that the interference effects in a system
with more than one interface can result in the formation of resonant
interface state. The most transparent manifestation of these states is
the resonant tunneling through a single-barrier structure. We shall
demonstrate this effect for the one-band effective mass model.

Let us consider a single-barrier heterostructure with a rectangular
potential barrier of the height $V$ and the width $L$. The
standard scattering solution to the Schr\"odinger equation is defined by
the following asymptotic form of the wave function
\begin{equation}
\label{Psis}
\begin{array}{c}{}\\ \psi(z)\\{}\end{array}=
\left\{ \begin{array}{lccl} e^{ikz}&+&r_{k}e^{-ikz},& 
\qquad z\rightarrow -\infty\\
t_{k}e^{ikz},&{}&{}&\qquad z\rightarrow +\infty \end{array} \right.
\end{equation}
where the wave vector $k$ is related to the energy of the incident
wave $E=k^{2}/2m$. Using the general connection rules Eq.~(\ref{BC})
and performing straightforward calculations we arrive at the following
expression for the transparency coefficient $D_{k}=|t_{k}|^{2}$
\begin{eqnarray}\nonumber
 &D_{k}& = 4k^2\kappa^2
\left\{ \left[
 \left(\kappa^2T_{11}^2 + k^2T_{22}^2 + T_{21}^2 +
 k^2\kappa^2T_{12}^2\right)\sinh \kappa L    
\right. \right.\\
&+&\left. \left. 
 2\kappa\left(T_{11}T_{21} + k^2T_{22}T_{12}\right)
\cosh\kappa L \right]^2 + 4k^2\kappa^2 \right\}^{-1}
\label{D}
\end{eqnarray}
where
$$
\kappa=\sqrt{2mV-k^2}.
$$
The transparency $D_{k}$ Eq.~(\ref{D}) has an explicit resonant
structure and turns into unity at the condition
\begin{equation}
\label{resC1}
\tanh \kappa L= -\frac{2\kappa\left(T_{11}T_{21}+
k^2T_{22}T_{12}\right)}{\kappa^2T_{11}^2+
k^2T_{22}^2+T_{21}^2+k^2\kappa^2T_{12}^2}.
\end{equation}
If $L\rightarrow \infty$, Eq.~(\ref{resC1}) has no solutions. However,
the resonance occurs if the width of the barrier $L$ becomes
smaller than some critical value $L_{c}$ and if at least one of the
off-diagonal elements is negative. 

To reveal the physical nature of the resonance condition
Eq.~(\ref{resC1}) we consider the simplest nontrivial transfer matrix with
only one nonzero off-diagonal element. Namely, we assume that
$T_{11}=T_{22}=1$, $T_{12}=0$ and $T_{21}\ne 0$. This transfer
matrix corresponds to the $\delta$-function interface potential of the
strength $g=2mT_{21}$. Since the resonant solution exists only for
negative $T_{21}$ we also assume that $T_{21}<0$. Under the above
assumptions the resonance condition Eq.~(\ref{resC1}) reduces to the equation
\begin{equation}
\tanh\kappa L =\frac{2\kappa|T_{21}|}{2mV + T_{21}^{2}}.
\label{resC2} 
\end{equation}
Introducing the following dimensionless variables
\begin{eqnarray}
\label{x}
x &=& \frac{\kappa}{\sqrt{2mV}}=\sqrt{1 - \frac{E}{V}},\\
\label{tau}
\tau &=& 2\frac{|T_{21}|/\sqrt{2mV}}{1 + 
\left(|T_{21}|/\sqrt{2mV}\right)^{2}},\\
\label{l}
l &=& L\sqrt{2mV},
\end{eqnarray}
we transform Eq.~(\ref{resC2}) to the form
\begin{equation}
\tanh xl = \tau x.
\label{resC3}
\end{equation}
Parameter $\tau$ in the right hand side in Eq.~(\ref{resC3}) can not
exceed unity ($\tau\le 1$). In fact, the function
$\tau(|T_{21}|/\sqrt{2mV})$ Eq.~(\ref{tau}) (see Fig.~4) reaches its
maximum value $\tau_{\max}=1$ at
\begin{equation}
|T_{21}|/\sqrt{2mV}=1.
\label{max}
\end{equation}
\begin{figure}
  \includegraphics[width=0.4\textwidth]{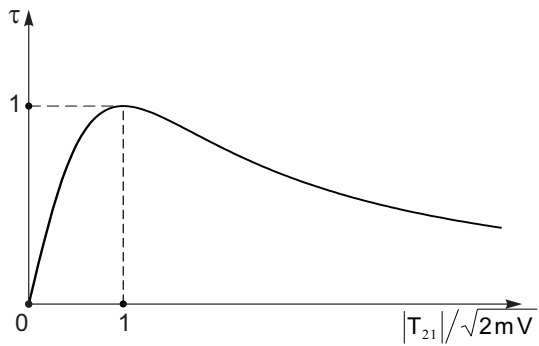}
  \caption{Functional dependence of $\tau$ on
    $|T_{21}|/\sqrt{2mV}$ (Eq.~(\ref{tau})).}
  \label{fig4}
\end{figure}
Note, that Eq.~(\ref{max}) defines  the critical value of
$|T_{21}|$ which is required for the existence of the localized state
at the single heterointerface. Indeed, Eq.~(\ref{23}) shows that the
interface bound state exists if $|T_{21}|/\sqrt{2mV}>1$. Since the
behavior of solutions to Eq.~(\ref{resC3}) is governed by the
parameter $\tau$, the existence of the under-the-barrier resonances
does not depend on 
whether or not the separate single interface possesses bound states. We
shall see, however, that these resonances are closely
related to the bound states which exist in the combined double-interface
structure.   

Under-the-barrier resonances correspond to the solutions to
Eq.~(\ref{resC3}) in the region $0<x<1$ ($0<E<V$). Graphical solution
of Eq.~(\ref{resC3}) is illustrated in Fig.~5 where the functions
$\tau x$ and $\tanh xl$ are plotted by dashed and solid lines
respectively. 
The required solution exists if parameter $l$, which is
proportional to the width of the barrier, satisfies the inequalities
\begin{equation}
\tau < l < l_{c},
\label{rangel}
\end{equation}
where $l_{c}$ is the solution to the equation $\tanh l_{c}=\tau$
\begin{equation}
l_{c} = \frac{1}{2}\ln\frac{1+\tau}{1-\tau}.
\label{l_{c}}
\end{equation}
Figure~5 shows that with decrease of $l$ we first meet the resonance
condition at $l=l_{c}$. The solution $x=1$ corresponds to the
resonant transparency $D_{k}=1$ at zero energy $E=0$. With the further
decrease of $l$ the resonance energy moves up, reaches the the top of
the barrier $E=V$ ($x=0$) at $l=\tau$, and enters the over-the-barrier
continuum. 
\begin{figure}
  \includegraphics[width=0.4\textwidth]{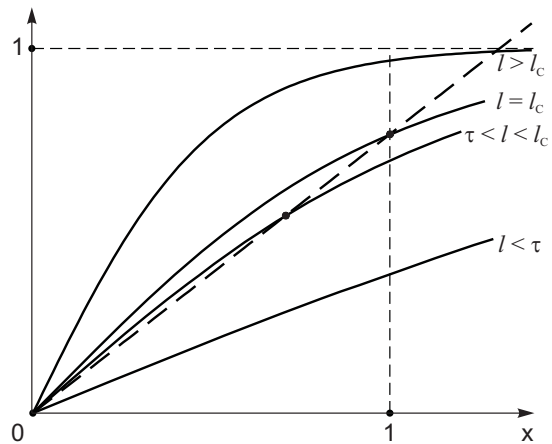}
  \caption{Graphical solution of the equation
    $\tanh xl=\tau x$ (Eq.~(\ref{resC3})) which defines the behavior
    of the under-the-barrier resonance.}
  \label{fig5}
\end{figure}

The asymptotic form of the scattering solution Eq.~(\ref{Psis}) at $k=-iq$,
$t_{iq}=\pm 1$ and $r_{iq}=0$ exactly coincides with the asymptotic form
of a localized solution. Hence the condition for the resonant
transparency at zero energy $|t_{k=0}|=1$ is, in fact, the condition
for appearance/disappearance of a localized state (symmetric or
antisymmetric). Thus, the resonant state is nothing but the localized
state (obviously antisymmetric) which is pushed out of the band gap.   

Now we are able to describe the evolution of both resonant and
localized states with the change of the barrier width $L$ at fixed $V$
and $|T_{21}|$. We consider separately two different cases:

(i) $|T_{21}|/\sqrt{2mV}>1$. At $L\rightarrow \infty$ there exist two
degenerate localized states which are related to the well
separated heterointerfaces. With decrease of $L$  the degeneracy is
lifted and, at $L=L_{c}=l_{c}/\sqrt{2mV}$, the upper state is pushed
out of the band gap to form the resonance. With the further decrease of $L$
the resonance moves up and crosses the top of the barrier at
$L=\tau\sqrt{2mV}$. At $L<\tau\sqrt{2mV}$ we have no resonance, but
only the localized state.     

(ii) $|T_{21}|/\sqrt{2mV}<1$. At $L>L_{c}$ there exists neither localized
nor resonant state. The critical value $L=L_{c}$ corresponds to the
creation of the resonance-bound state pair. When $L$ becomes smaller
than $L_{c}$ the resonance moves up and reaches the top of the barrier
at $L=\tau\sqrt{2mV}$, whereas the localized state moves down into the
energy gap. 

Thus in either case the resonant transparency is always
accompanied by the bound state. The resonance and the localized state
can be qualitatively interpreted as anti-bonding and bonding orbitals
respectively. 

It is interesting to note that in the simple case of zero element $T_{12}$
the resonant transparency of the single-barrier structure has a clear
counterpart in the classical electrodynamics. The corresponding system
is a metallic slab which is covered on either side by dielectric
layers. The metallic region with $\varepsilon(\omega)<0$ models the
barrier, whereas the dielectric coating corresponds to the attractive
interface potential. The resonant transparency of electromagnetic
waves through such a system has been described
theoretically\cite{Zharov,Dragila} and observed experimentally in
Ref.~\onlinecite{Dragila}. If both off-diagonal elements are nonzero,
the resononant tunneling through a single-barrier heterostructure
apparently has no an optical analog. 

\section{Conclusion}

It it commonly accepted that effects of a microscopic structure of a
heterointerface can be incorporated into the envelope function method
by the use of generalized connection rules. Such connection rules
are normally formulated in terms of the interface transfer matrix. In
this paper we presented the general method which allows to construct
the transfer matrix for arbitrary system. We showed that the Hermiticity
of the Hamiltonian and the symmetry of the interface plane impose the
restrictions on the form of transfer matrix. These restrictions can be
formulated as a set of equations for components of the transfer
matrix. Solution of these equations defines the general form of the
boundary conditions for a given interface. To illustrate this approach
we considered the one-band and 4-band envelope function models and
established the conditions for the existence of Tamm-like localized
interface states. We have also demonstrated that in a system with more than
one heterointerface there exists a possibility for new physical
effects such as the resonant tunneling through a single potential barrier.

In this paper we did not discuss the application of our general
results to particular heterostructures. Such an application requires a
knowledge of phenomenological parameters which enter the generalized
boundary conditions. Identification of these parameters using the
results of experiments or {\it ab initio} calculations is by far not
simple task. We believe, however, that the physical effects which have
been described in our paper and which originate solely from the
nontrivial structure of the boundary conditions, could be helpful in
resolving this problem. 

\section{Acknowledgment}

This work has been supported by Russian Federal Program
"Integratsiya" and Russian Program "Physics of Solid State
Nanostructures". I.~T. is grateful to the Alexander von Humboldt
Foundation for support.


\end{document}